\documentclass[twocolumn,preprintnumbers,amsmath,amssymb]{revtex4}
\usepackage{graphicx}
\usepackage{dcolumn}
\usepackage{bm}
\usepackage{color}

\begin{document}

\title{Dynamical local and non-local Casimir atomic phases}

\author{Fran\c{c}ois Impens$^{1,2}$, Claudio Ccapa Ttira$^{2}$,  Ryan O. Behunin$^{3,4,5}$, and Paulo A. Maia Neto$^{2}$}

\date{\today}

\affiliation{$^{1}$ Observatoire de la C\^{o}te d'Azur (ARTEMIS), Universit\'e de Nice-Sophia Antipolis, CNRS, 06304 Nice, France} 
\affiliation{$^{2}$ Instituto de F\'{i}sica, Universidade Federal do Rio de Janeiro,  Rio de Janeiro, RJ 21941-972, Brazil }
\affiliation{$^{3}$ Theoretical Division, MS B213, Los Alamos National Laboratory, Los Alamos, NM 87545, USA}
\affiliation{$^{4}$  Center for Nonlinear Studies, Los Alamos National Laboratory, Los Alamos, New Mexico 87545, USA}
\affiliation{$^{5}$  Department of Applied Physics, Yale University, New Haven, Connecticut 06511, USA}

\begin{abstract} 
We develop an open-system dynamical theory of the Casimir interaction between coherent atomic waves and a material surface. The system --- the external atomic waves --- disturbs the environment --- the electromagnetic field and the atomic dipole degrees of freedom --- in a non- local manner by leaving footprints on distinct paths of the atom interferometer. This induces a non-local dynamical phase depending simultaneously on two distinct paths, beyond usual atom-optics methods, and comparable to the local dynamical phase corrections. Non-local and local atomic phase coherences are thus equally important to capture the interplay between the external atomic motion and the Casimir interaction. Such dynamical phases are obtained for finite-width wavepackets by  developing a diagrammatic expansion of the disturbed environment quantum state. 

\end{abstract}


\date{\today}

\maketitle

\section{INTRODUCTION}

The interplay between the internal atomic dynamics and the electromagnetic (EM) field retardation, brought to light by the pioneering work of 
Casimir and Polder~\cite{CasimirPolder}, is crucial to  understand  the atom-surface dispersive interaction in the long-distance limit (see \cite{Intravaia} for a recent review).
 In contrast, the 
 effect of the external atomic motion on the dispersive interaction
 is almost always discarded. 
 Notable exceptions are the quantum friction effects resulting from the shear relative motion between two material surfaces ~\cite{QuantumFrictionPP}
  or between an atom and a surface \cite{QuantumFrictionAP,Scheel09}. 

Since the usual atomic velocities are strongly non-relativistic, 
one might expect  the dynamical corrections  to the dispersive atom-surface interaction to be very small.
Because of their high sensitivity, atom interferometers \cite{Cronin09,Kasevich07} are ideal systems for probing such small corrections. 
  There is a growing interest in developing atom interferometers able to  probe surface interactions. Measurements of the van der Waals atom-surface
   interaction with standard atom interferometry have already been achieved~\cite{Cronin04,CroninVigue,Lepoutre11}, while
   optical-lattice atom interferometry offers even more promising perspectives to measure the Casimir-Polder interaction in the 
   long-distance regime~\cite{FORCAGpapers}.

From a fundamental point of view, the coherent atomic waves evolving in the vicinity of a material
 surface constitute a particularly rich open quantum system: the external atomic waves, playing the role of the system, interact with an environment involving both long-lived (atomic dipole) and short-lived (EM field) degrees of freedom (dofs).
  In this paper, we develop an open-system theory of atom interferometers in the vicinity of a material surface. 
  We show that the atomic motion relative to the surface along 
the interferometer paths  gives rise to a non-local dynamical phase correction associated to 
  pairs of paths rather to individual ones as in usual interferometers. 
   In contrast to the local dynamical phase contributions, the non-local dynamical phases may be distinguished from other quasi-static phase contributions
   in a multiple-path atom interferometer~\cite{MultiplePathAtomInterferometer} since they violate additivity~\cite{NonAdditiveCasimir}.

  Preliminary results for extremely narrow wavepackets  were
  derived in a previous letter ~\cite{DoublePath}
  from the influence functional~\cite{FeynmanVernon} 
   capturing the net effect of the environment on the atomic center of mass (external) dynamics~\cite{Ryan10,Ryan11}.
  The atomic phases were then calculated in terms of
 closed-time path integrals~\cite{CalzettaHu}.
 
  Here we use instead  standard perturbation theory to investigate the more realistic case of finite-width wavepackets, 
  allowing us to connect with the van der Waals interferometer experiments~\cite{CroninVigue}.  
   We explicitly calculate the disturbance of the environment \cite{SAI90} 
    produced by the interaction with the external dofs in the atom interferometer. 
    Since the perturbation  is of second-order, the changes of the environment state involves two atomic ``footprints'', which can be left either on the same path, or on distinct paths.
    Provided that the dipole memory time is longer than the time it takes for light to propagate between the two arms,
    the diagrams for which the atomic waves have ``one foot on each path'' yield cross non-local phase contributions. For atoms flying parallel to the plate, these cross contributions cancel each other exactly. 
    Otherwise, the differential atomic motion between the two interferometer arms  brings into play an asymmetry between the cross-talk diagrams,
    thanks to 
    the finite velocity of light and the breaking of the translational invariance by the surface.    
  The resulting  non-local  phase contribution is of the same order of magnitude
    of the dynamical local corrections. 
 Non-local phase coherences
     are thus required
     in a consistent description of 
       dynamical  effects in Casimir atom interferometry.

  Our formalism also allows for the analysis of the decoherence effect in interferometers~\cite{Barone,Hackermueller04,Breuer01,Lamine06} 
  in the presence of a conducting plane~\cite{CasimirDecoherence,Sonnentag07}.
The analysis of the
   path-dependent disturbance of the environment  provides  a clear-cut approach to the derivation of  decoherence~\cite{SAI90},
   which was employed 
in the derivation of the dynamical Casimir decoherence for neutral macroscopic bodies \cite{Dalvit00}. 
 Alternatively, the decoherence 
 effect can be obtained from 
 the modulus of the complex influence functional~\cite{Mazzitelli03}, which depends on the imaginary part of the environment-induced phase shift. 
However, here we focus on the 
 real part of the 
 Casimir phase shift, which has been measured experimentally for neutral atoms \cite{CroninVigue}, in contrast with the  loss of contrast in the fringe pattern,
 which has been probed only in the case of 
 charged particles \cite{Sonnentag07}. 
 Environment-induced phase shifts were also considered in the context of geometrical phases for spin one-half systems \cite{GeometricPhase}.

We shall proceed as follows. In Sec.~\ref{sec:local dynamical Casimir phases}, we develop a local dynamical theory of Casimir atom interferometers, inspired by the atom-optical $ABCD$ formalism~\cite{BordeABCD},  and show its consistency with the standard phase obtained from the
dispersive potential in the quasi-static limit. 
In the following sections, we go beyond this heuristic treatment by considering the disturbance of the environment 
quantum state by the interaction with the external atomic waves, first in the simpler case of 
point-like wave-packets in  ~\ref{sec:Feynman diagrams} and then for finite-width wave-packets in \ref{sec:finite width}.
This treatment reveals the appearance of dynamical non-local atomic phase coherences in addition to the local contributions already obtained in Sec.~\ref{sec:local dynamical Casimir phases}. 
Explicit results for the case of a perfectly-reflecting plane surface are derived in Sec.~\ref{section:relativistic expansion}
and concluding remarks are presented in Sec.~\ref{section:conclusion}.

\section{LOCAL DYNAMICAL THEORY OF CASIMIR PHASES}
\label{sec:local dynamical Casimir phases}

\begin{figure}[htbp]
 \begin{center}
 \includegraphics[width=8.5cm]{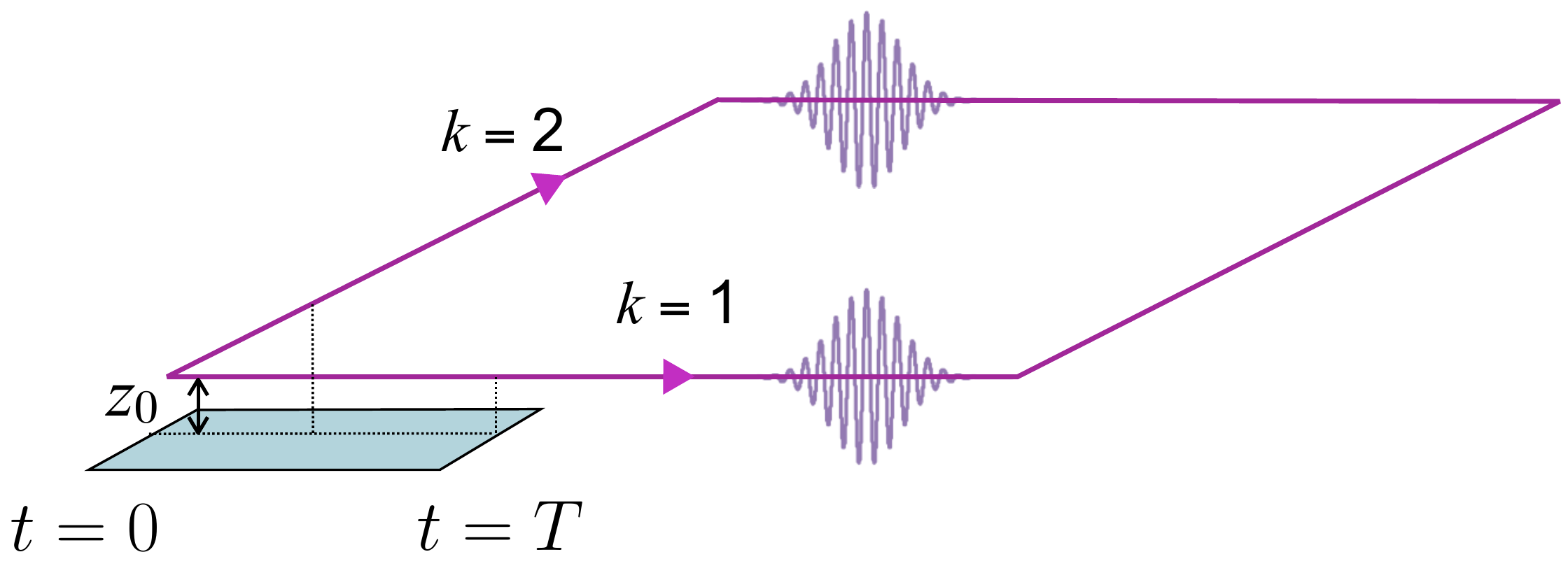}
 \end{center}  \caption{(color
   online).
 Atom interferometer 
 interacting with a 
  conducting plate at $z=0$ during the time $T,$ with the arm $k=1$ parallel to the plate (distance $z_0$) 
   and the arm $k=2$ flying away with a normal velocity $v_{\perp}.$ 
    }
 \label{fig:atom interferometer}
 \end{figure}

 In this section, we develop  a local theory of a Mach-Zehnder atom interferometer in interaction with a material surface (see Fig.~\ref{fig:atom interferometer} for a typical example). In contrast to the idealized point-like model discussed in Ref.\cite{DoublePath}, the derivation
  below fully captures the influence of the wave-packet finite width, making our discussion relevant for atom interferometers with large wave-packets, such as those employed in the recent experiments reported in Refs.\cite{CroninVigue}.
 
 In the usual closed-system approach, the atom-surface interaction phase is given by the  integration of an external dispersive potential taken at the instantaneous atomic position. Obviously, this standard approach is completely quasi-static -- the potential seen by the atoms depends only on their 
  instantaneous position distribution, but not on their velocity. Here, we perform instead a first-principle derivation of this phase based on the interaction energy stored within the quantum dipole and EM field dofs. While capturing non-trivial local relativistic corrections,
   this treatment yields predictions in agreement with the standard dispersive potential approach when considering the quasi-static limit.

 The atomic wave-function is initially a coherent superposition $| \psi_E (0) \rangle = \frac {1} {\sqrt{2}} \left(  |\psi_{E}^1(0) \rangle + | \psi_{E}^2(0)  \rangle \frac {} {}  \right) $ of two wave-packets with the same central position but with different initial momenta. These wave-packets will follow two distinct paths $k=1,2$ as illustrated in Fig.~\ref{fig:atom interferometer}. The relative phase between these two wave-packets, which determines the local atomic probability function $p(\mathbf{r},t)=|\psi_E(\mathbf{r},t)|^2$, contains contributions from the atom-surface interaction as well as additional
 ones independent of the surface.

 As in Ref.~\cite{DoublePath}, we extend the atom-optics $ABCD$ formalism~\cite{BordeHouches,BordeABCD,AtomLaserABCD} by including the symmetrized~\cite{DalibardRocCohen} interaction energy $U^{\rm int}_k(t)$  between the atomic dipole and the EM field within the action phase associated to the external atomic propagation along path $k.$  The atom-surface interaction, assumed weak enough to leave unaltered the shape of the atomic wave-packets during the propagation, results merely in  atomic phase shifts.
 
 We evaluate $U^{\rm int}_k(t)$ using linear response theory~\cite{WylieSipe}, i.e to lowest order in perturbation theory, and then 
  obtain the local Casimir phase $\varphi^{\rm loc}_k= - \frac {1} {\hbar} \int_0^T dt U_k^{\rm int,S}(t)$ along a given path $k$ 
  by picking the surface-dependent contribution $ U_k^{\rm int,S}(t)$ to the total interaction energy. 
  The key ingredient in our derivation  is 
     the introduction of an ``on-atom field'' operator $\hat{\mathbf{E}}(\hat{\mathbf{r}}_a),$ for which the field argument is the
      atomic position operator $\hat{\mathbf{r}}_a$ instead of a classical position $\mathbf{r}_k(t)$ taken along the central atomic path $k$. 
     
     In the Heisenberg picture, the dipole and the on-atom electric field operators can be expressed as the sum of an unperturbed free-evolving part, 
     defined as $\hat{\mathbf{O}}^f(t)= \exp \left( i \hat{H}_0 t/\hbar \right) \hat{\mathbf{O}}(0) \exp \left( -i \hat{H}_0 t/\hbar \right) $ 
     with the free Hamiltonian $\hat{H}_0=\hat{H}_E+\hat{H}_D+\hat{H}_F$
      including the external ($H_E$), internal ($H_D$) and EM field ($H_F$) dofs,
      and of a contribution $\hat{\mathbf{O}}^{in}(t)$ induced by the atom-field coupling $\hat{H}_{AF} = - \hat{\mathbf{d}} \cdot \hat{\mathbf{E}} (\hat{\mathbf{r}}_a).$ To describe the mutual influence between the atomic dipole and the `on-atom' EM
  field~\cite{WylieSipe}, we introduce temporal correlation functions for the corresponding operators. We also introduce four-point correlation functions for the quantized electric field as discussed below.
  
 Precisely, the dipole and field fluctuations are captured by symmetric correlation functions (also refered to as Hadamard Green's functions) 
 of the free-evolving operators $\hat{\mathbf{O}}^f=\hat{\mathbf{d}}^f(t), \hat{\mathbf{E}}(\hat{\mathbf{r}}_a)^f(t), \hat{\mathbf{E}}^f(\mathbf{r},t)$
  ($\{...\}$ denotes the anti-commutator):
  \begin{eqnarray}
  \label{eq:Hadamard Green functions}
   G^{H}_{\hat{\mathbf{O}}, \: ij}(x;x')  =   \frac 1 \hbar \langle  \{ \hat{O}_i^{f}(x),  \hat{O}_j^{f}(x')  \} \rangle. \label{Hadamard}
  \end{eqnarray}
   For the dipole and on-atom field operators
  $\hat{\mathbf{O}}=\hat{\mathbf{d}}, \hat{\mathbf{E}}(\hat{\mathbf{r}}_a)$ 
  the arguments in (\ref{Hadamard}) are two instants $(x;x') \equiv (t,t')$. For the electric field operator $\hat{\mathbf{O}}=\hat{\mathbf{E}},$ 
  these arguments are two four-vectors $(x;x') \equiv (\mathbf{r},t;\mathbf{r}',t').$  
  
    The linear susceptibilities (polarizability for the dipole), generically written as 
    retarded Green's functions, describe the linear response of 
  field and dipole to dipole and field
   perturbations, respectively:
  \begin{eqnarray}
  \label{eq:retarded Green functions}
  G^{R}_{\hat{\mathbf{O}}, \: ij}(x;x')  = \frac {i} {\hbar} \theta(t-t') \langle [ \hat{O}_i^{f}(x),  \hat{O}_j^{f}(x')  ] \rangle
  \end{eqnarray}
    with $\theta(t-t') $ denoting the Heaviside step function.
  
  Note that the on-atom field Green's functions as defined by 
  (\ref{eq:Hadamard Green functions}) and (\ref{eq:retarded Green functions}) are still quantum operators in the Hilbert space corresponding to the 
  atomic external dofs, since the average is taken over the EM field dofs only. We now take the average
   $\langle \: \mathcal{G}_{\hat{\mathbf{E}}(\hat{\mathbf{r}}_a)}^{R,H}(t,t') \: \rangle_k $
   over the external quantum state $ | \psi^k_{E} \rangle $ 
    corresponding to the single atomic wave-packet  $k$.  
We express the result in 
  terms of the atomic wave-functions  $\psi_E^k(\mathbf{r},t)=\langle \mathbf{r} | e^{ - \frac i \hbar \hat{H}_E t} | \psi_E^k(0) \rangle, $ of the external atomic  propagator
   \begin{equation}
   \label{eq:definition atomic propagator}
    K(\mathbf{r},t;\mathbf{r}',t')= \langle \mathbf{r} | e^{- \frac {i} {\hbar}  H_E (t-t')} | \mathbf{r}' \rangle,
    \end{equation}
 and of the electric field Green's functions. For this purpose, we switch to the Schr\"odinger picture with respect to the external atomic dofs: $\hat{\mathbf{E}}(\hat{\mathbf{r}}_a)(t) = e^{\frac {i} {\hbar} H_E t}  \hat{\mathbf{E}}(\hat{\mathbf{r}},t)   e^{-\frac {i} {\hbar} H_E t}  $ with $\hat{\mathbf{r}} = \hat{\mathbf{r}}_a(0)$ the atomic position operator, and $\hat{\mathbf{E}}(\mathbf{r},t)$ the quantized electric field (Heisenberg-evolved with respect to the Hamiltonian $H_F$) at the classical position $\mathbf{r}$ and  time $t$.  Using closure relations for the external atomic dofs, one obtains
   \begin{eqnarray}
   \label{eq:on atom Green function wide atomic wave packets}
   \langle \mathcal{G}_{\hat{\mathbf{E}}(\hat{\mathbf{r}}_a)}^{R(H)}(t',t) \rangle_k & = &  \! \iint \! d^3\mathbf{r} d^3\mathbf{r}' 
    \psi_E^{k *}(\mathbf{r},t) K(\mathbf{r},t;\mathbf{r}',t') \psi_E^k(\mathbf{r}',t') \nonumber \\
   & & \qquad \qquad  \times \mathcal{G}_{\hat{\mathbf{E}}}^{R(H)}(\mathbf{r},t;\mathbf{r}',t') \, .
   \end{eqnarray}
   
     It is necessary to identify the physically relevant contributions of the field response (and fluctuations) as far as the atom-surface interaction is concerned.  By isotropy of the atomic dipole, only the trace of the  electric field
                 Green's functions 
                  $\mathcal{G}^{R (H)}_{\hat{\mathbf{E}}}(x;x') \equiv \sum_i G^{R (H)}_{\hat{\mathbf{E}} \: i i}(x;x')$ (with the sum performed on the Cartesian index $i=1,2,3$) is needed to obtain the interaction energy. $\mathcal{G}^{R (H)}_{\hat{\mathbf{E}}}(x;x')$ is the sum of free-space and scattering contributions:
               \begin{equation}\label{0S}
               \mathcal{G}^{R (H)}_{\hat{\mathbf{E}}}(x;x')  = \mathcal{G}^{R (H),0}_{\hat{\mathbf{E}}}(x;x') +
               \mathcal{G}^{R (H),S}_{\hat{\mathbf{E}}}(x;x') 
               \end{equation}
               By symmetry the free-space contributions  $ \mathcal{G}^{R (H),0}_{\hat{\mathbf{E}}}({\bf r},t;{\bf r}',t') $ depends only 
               on $|{\bf r}-{\bf r}'|$ and $t-t'$ \cite{Heitler}, whereas 
               the scattering contribution $ \mathcal{G}^{R (H),S}_{\hat{\mathbf{E}}}({\bf r},t;{\bf r}',t')$  can be written in terms of 
               the image of the source point ${\bf r}'$ in the particular case of a planar perfectly-reflecting surface discussed in Sec. IV. 
               More specifically, the free-space retarded Green's function $\mathcal{G}^{R,0}_{\hat{\mathbf{E}}}({\bf r},t;{\bf r}',t') $ represents
               the direct propagation from ${\bf r}'$ to $\bf r$ and does not depend on the distance to the material surface, whereas 
               the scattering contribution  $\mathcal{G}^{R,S}_{\hat{\mathbf{E}}}({\bf r},t;{\bf r}',t') $ corresponds to the propagation with one reflection at the surface. 
               
               When replacing  (\ref{0S}) into 
               (\ref{eq:on atom Green function wide atomic wave packets}), the average on-atom  field Green's functions also split into
               free-space and scattering contributions, and 
               only the latter
               contributes to the atom-surface interaction energy $U_k^{\rm int,S}(t)$ and hence to the local Casimir phase $\varphi^{\rm loc}_k.$
   The latter is derived by following steps similar to those employed for point-like wave-packets and using  expression (\ref{eq:on atom Green function wide atomic wave packets}) with the field Green's function replaced by the scattering contribution 
    $\mathcal{G}^{R (H),S}_{\hat{\mathbf{E}}}({\bf r},t;{\bf r}',t'):$               
    \begin{eqnarray}
 \label{eq:local phase general wide atomic packets}
 \varphi^{\rm loc}_k \!& = & \!\frac {1} {4} \! \iint_{0}^{T} \! d t  dt' \! \iint \! d^3\mathbf{r} d^3\mathbf{r}' \! \psi_E^{k *}(\mathbf{r},t) \! K(\mathbf{r},t;\mathbf{r}',t')\! \psi_E^k(\mathbf{r}',t') \nonumber \\
 & \times & \left[  g_{\hat{d}}^H(t,t') \mathcal{G}_{\hat{\mathbf{E}}}^{R,S}(\mathbf{r},t;\mathbf{r}',t')
  +    g_{\hat{d}}^R(t,t')  \:   \mathcal{G}_{\hat{\mathbf{E}}}^{H,S}(\mathbf{r},t;\mathbf{r}',t')  \right]. \nonumber \\
 \end{eqnarray}
 with $  g_{\hat{d}}^{R(H)}(t,t') $ representing any diagonal component of the isotropic atomic dipole Green's function $G^{R(H)}_{\hat{\mathbf{d}}, \: ii}(t,t').$
  The two contributions appearing in (\ref{eq:local phase general wide atomic packets}) correspond to the separate physical effects responsible for the 
 atom-surface dispersive interaction: radiation reaction and field fluctuations~\cite{Meschede90,Mendes}. The former, proportional to the field retarded Green's function, 
 dominates in the van der Waals  un-retarded short-distance limit and is of particular relevance in the following sections. Physically, it represents the self-interaction 
 between the fluctuating dipole at time $t$ and position $\bf r$ with 
 its own electric field, produced at an earlier time $t'$ and position ${\bf r}',$ after  
 bouncing off the material surface.  This interpretation provides an indication that  a cross non-local interaction might also exist, with the field produced at one wave-packet component propagating to a different wave-packet component, as discussed in detail in the following sections.

 As a first check of (\ref{eq:local phase general wide atomic packets}), we consider the limit of very narrow wave-packets in order to compare 
 with  Ref.~\cite{DoublePath}.
We assume that the wave-packet width is much shorter than the relevant EM field wave-lengths, 
and then approximate the  position arguments of the Green's functions $\mathcal{G}_{\hat{\mathbf{E}}}^{(R)H,S}(\mathbf{r},t;\mathbf{r}',t')$ by the central atomic positions $\mathbf{r}_k(t)$ and $\mathbf{r}_k(t')$ taken along the trajectory $k$ at the respective times $t,t'.$
In this case, we can isolate the
 atomic propagation integral $\psi_E^k(\mathbf{r},t)= \int d^3 \mathbf{r}' K(\mathbf{r},t;\mathbf{r}',t')\! \psi_E^k(\mathbf{r}',t')$ in 
  (\ref{eq:local phase general wide atomic packets}) and find 
  \begin{eqnarray}
\label{eq:local phase general expression narrow wavepackets}
\varphi^{\rm loc}_k \! & \approx & \!\frac {1} {4} \! \iint_{0}^{T} \! d t  dt' \! 
\left[ \frac {} {}  g_{\hat{d}}^H(t,t') \mathcal{G}_{\hat{\mathbf{E}}}^{R,S}(r_k(t) , r_k(t') ) \right.  \\
& \: & \left.  \qquad \qquad  \qquad \qquad + g_{\hat{d}}^{R}(t,t')  \:   \mathcal{G}_{\hat{\mathbf{E}}}^{H,S} (r_k(t),r_k(t') )  \frac {} {} \right].  \nonumber
\end{eqnarray} 
in agreement with 
 Ref.~\cite{DoublePath}.
 
 A second, more important limiting case of 
 Eq.~(\ref{eq:local phase general wide atomic packets}), 
 corresponds to its quasi-static limit. 
 We also assume thermal equilibrium for the dipole and EM field dofs, and consider long interaction times (stationary regime). 
 In this case, the dipole and electric field Green's functions depend only on the time difference $\tau=t-t'$ and not on the individual times. 
 The retarded Green's functions $\mathcal{G}_{\hat{\mathbf{E}}}^{R,S}(\mathbf{r},\tau;\mathbf{r}', 0)$ is non-zero
only for a time delay $\tau$  equal to the time it takes for a photon to travel from the source position $\mathbf{r}'$ to the position $\mathbf{r}$
after one reflection at the surface.  
These durations are, in usual experimental conditions, much shorter than the time scales associated with the external atomic motion. In the quasi-static limit,
 we treat the external atomic motion as completely ``frozen'' during the time delay $\tau=t-t'$. In other words, 
we take $t' := t$  in the
external atomic propagator and wave-functions. 
In this limit, the former simplifies to $K(\mathbf{r},t;\mathbf{r}',t) = \delta (\mathbf{r}-\mathbf{r}')$. 
The resulting expression  can be directly compared with the formula for the dispersive atom-surface potential
 $V_{\rm Cas}({\bf r})$~\cite{WylieSipe} as detailed in the Appendix.  We then  find that the local phase becomes a time integral of the dispersive potential 
 taken at the instantaneous atomic position weighted by the external probability density:
  \begin{equation}
  \label{eq:nonrelativstic limit 5 wide}
  \varphi^{\rm    loc }_k \approx - \frac {1} {\hbar} \int_{0}^{T} d t   \int d^3\mathbf{r}   \: |\psi_E^k(\mathbf{r},t)|^2  \: V_{\rm{Cas}}(\mathbf{r}). 
  \end{equation}
 
  The quasi-static expression (\ref {eq:nonrelativstic limit 5 wide}) was employed as the theoretical model for comparison 
  with experiments \cite{Cronin04,CroninVigue,Lepoutre11}. On the other hand,
  our more general result (\ref{eq:local phase general wide atomic packets}) allows for non-equilibrium~\cite{Ryan11, Antezza} and non-stationary regimes which 
 cannot be described by the more standard  expression (\ref{eq:nonrelativstic limit 5 wide}). 
Explicit results for the dynamical corrections to order ${\bf \dot r}_k(t)/c$ were derived in  Ref.~\cite{NonAdditiveCasimir}
 in the  case of very narrow atomic packets flying close to a perfectly-reflecting planar surface.  Note,
 however,  that we also find non-local atomic phase corrections  to order ${\bf \dot r}_k(t)/c.$
   Thus, a full quantum open system approach, to be developed in the next sections, is required to assess the first-order dynamical correction in a consistent way.

\section{NON-LOCAL DYNAMICAL CASIMIR ATOMIC PHASES}
\label{sec:Feynman diagrams}

From now on, we no longer model the effect of surface interactions as a local phase shift imprinted on each external atomic wave-packet. 
We consider instead the evolution of the full quantum state describing the external atomic waves, atomic dipole and EM field.  
In the discussion to follow, we will refer respectively to the dipole and EM field dofs as the ``environment'' 
and to the external atomic waves as the ``system''. We  consider the case of point-like wave-packets in this section, so as to introduce 
our method in a simpler setting, thus paving the way for the discussion of finite-width wave-packets in the following sections.

\begin{figure}[htbp]
\begin{center}
\includegraphics[width=8.5cm]{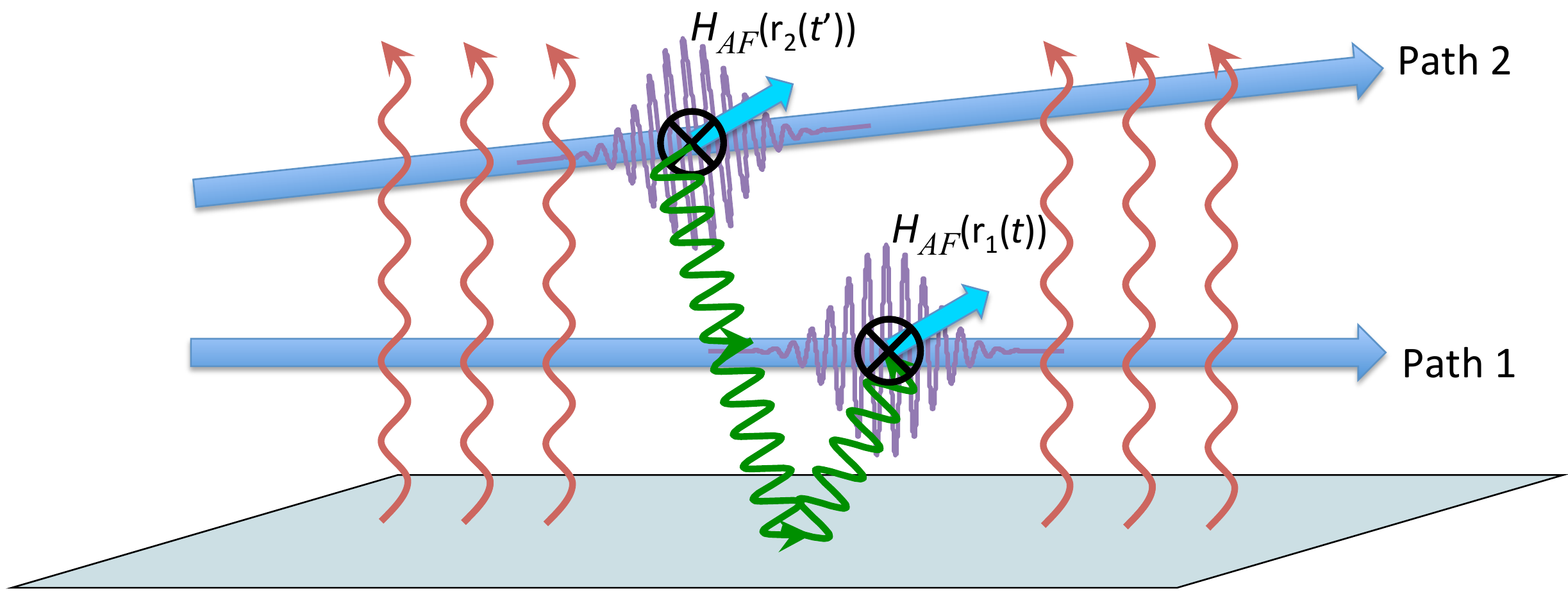}
\end{center}  \caption{(color
  online). Double-path footprint left on the environment (dipole + EM field) by the external atomic state through the dipolar interaction $\hat{H}_{AF}.$ 
   }
\label{new}
\end{figure}

We describe here how the quantum state of the environment is affected by the propagation of the external atomic waves.
Because it involves the center-of-mass position operator $\hat{\mathbf{r}}_a$, the dipolar Hamiltonian 
$\hat{H}_{AF}= - \hat{\mathbf{d}} \cdot \hat{\mathbf{E}} (\hat{\mathbf{r}}_a) $ 
operates on the environment in a manner which depends on the path followed by the atoms. Thus,
 such a Hamiltonian acts as a ``which-path'' marker, leaving an atomic ``footprint''  on the dipole and EM field quantum states. 
 The phase contribution  is of second order in the dipolar interaction Hamiltonian.
  A Feynman-diagram expansion shows that these footprints actually contain  cross terms, involving the two coherent components of the external atomic state propagating on two distinct arms of the interferometer
(see Fig.~\ref{new}).  
As discussed in detail below, such terms reflect a non-local disturbance of the environment operated at different times by the system. In addition to a loss of contrast in the fringe pattern, such perturbation also induces a non-local double-path atomic phase coherence. We derive here both the local and non-local phases resulting from the influence of the environment. The local phase shifts obtained below correspond exactly to the atom-surface interaction
 phases~
 \eqref{eq:local phase general wide atomic packets} and
 \eqref{eq:local phase general expression narrow wavepackets} derived in the previous section for finite-width and point-like wave-packets,
 respectively, whereas  the non-local phases cannot be derived from the interaction energy along the different paths taken 
separately.

In Ref.~\cite{DoublePath}, we have briefly outlined an alternative approach, based on the influence functional, which captures the 
effect of the environment on very narrow atomic waves as a complex phase which can also be recast as a stochastic phase~\cite{Ryan11}. 
This method leads to the same final results we derive in this section. The equivalence between the two
points of views illustrates  an important property of open systems~\cite{SAI90}: its evolution 
 is equally well described by considering the accumulation of a stochastic phase, or by analyzing the trace left by the system onto the quantum state of the environment.

\subsection{Atomic interferences in presence of an environment}

Inspired by Ref.~\cite{SAI90}, we calculate the time evolution of the full quantum state, which is initially given by
  $| \psi (0) \rangle = \frac {1} {\sqrt{2}} \left( | \psi_E^1(0) \rangle + | \psi_E^2(0) \rangle \right) \otimes | \Psi_{DF}(0) \rangle,$
  where 
$| \Psi_{DF}(0) \rangle= | \psi_D(0) \rangle \! \otimes \! | \psi_F(0) \rangle $ denotes the initial environment (internal dipole and EM field) quantum state.  
  We  discard the influence of the atom-surface interaction on the external atomic motion (prescribed atomic trajectories), 
 which is a very good approximation in usual experimental conditions~\cite{CroninVigue}. In this section, we
assume, for simplicity, that the  wave-packet width is much smaller than the relevant field wavelengths (more general results are derived in the following sections). 
 Thus, the interaction is described by the 
 Hamiltonians  $ \hat{H}_{AF}(\mathbf{r}_k(t))= - \hat{\mathbf{d}} \cdot \hat{\mathbf{E}} (\mathbf{r}_k(t)) $ 
 parametrized by the wave-packet trajectories represented by the four-vectors $r_k(t)\equiv (\mathbf{r}_k(t),t)$ 
 with $k=1,2$, and acting only on the dipole and EM field Hilbert spaces \cite{foot_Rontgen}. We
  work in the interaction picture  and  the transformed time-dependent interaction Hamiltonian reads
\begin{equation}
\label{eq:interaction Hamiltonian Heisenberg}
 \hat{\widetilde{H}}_{AF}(r_k(t)) =  e^{  \frac {i} {\hbar}  ( \hat{H}_D + \hat{H}_F) t }  \left(-\hat{\mathbf{d}} \cdot \hat{\mathbf{E}} (\mathbf{r}_k(t)) \right)  e^{  - \frac {i} {\hbar}   ( \hat{H}_D + \hat{H}_F) t }.
 \end{equation}
At time  $t=T$, the full quantum state reads
\begin{eqnarray}
| \psi(T) \rangle & =& \frac {1} {\sqrt{2}}  | \psi_E^1(T) \rangle \! \otimes \! \mathcal{T} e^{ - \frac {i} {\hbar} \int_0^T dt  \hat{\widetilde{H}}_{ AF}(r_1(t)) }   | \Psi_{DF}(0) \rangle
 \nonumber \\
& + & \frac {1} {\sqrt{2}}  | \psi_E^2(T)  \rangle \! \otimes \! \mathcal{T} e^{ - \frac {i} {\hbar} \int_0^T dt'  \hat{\widetilde{H}}_{ AF}(r_2(t))} \! | \Psi_{DF}(0) \rangle,\nonumber \\
\label{psi}
\end{eqnarray} 
where $\mathcal{T} $ denotes the  time-ordering operator.

 Since the dipole and EM field states are not measured in the experiment, 
 we calculate the external reduced density operator $\rho= {\rm Tr}_{DF} \left(| \psi(T) \rangle \langle  \psi(T) |\right).$
 When replacing  (\ref{psi}) into this equation,  the cross (interference) term represents the  
 external atomic coherence, which we evaluate in the 
position representation:
 \begin{equation}
 \label{eq:definition coherence rho12}
 \rho_{12}({\bf r},{\bf r}';T) = \frac12\,\langle  {\bf r}| \psi_E^1(T) \rangle \langle  \Psi_{DF}^2(T)|\Psi_{DF}^1(T)\rangle \langle \psi_E^2(T)  | {\bf r}' \rangle
 \end{equation}
Thus, 
the interference term
 $\rho_{12}^{(0)}=\frac12 \psi_E^2(\mathbf{r}',T)^*\,  \psi_E^1(\mathbf{r},T) $ is now
 multiplied by the scalar product  of the disturbed environment states
\begin{eqnarray}
\label{eq:environment quantum states product}
 \langle \Psi_{DF}^2(T) | \Psi_{DF}^1(T) \rangle  & \equiv &e^{i \Phi_{12}}.   
\end{eqnarray}
The complex phase  $\Phi_{12}$ captures 
 the environment effect on the external interference term accumulated over the interaction time $T$:
\begin{eqnarray}
\label{eq:environment quantum states product}
e^{i \Phi_{12}} &= &
  \langle \Psi_{DF}(0) |  \widetilde{\mathcal{T}} e^{  \frac {i} {\hbar} \int_0^T dt  \hat{\widetilde{H}}_{ AF}(r_2(t))}    \nonumber \\
 && \times  \mathcal{T} e^{ - \frac {i} {\hbar} \int_0^T dt  \hat{\widetilde{H}}_{ AF}(r_1(t))} \! | \Psi_{ DF}(0) \rangle
\end{eqnarray}
with $\widetilde{\mathcal{T}} $ denoting the anti-time-ordering operator (earlier-time operators on the left).

In general 
the final environmental quantum states have a scalar product smaller than unity $ |\langle \Psi^2_{DF }(T)  
| \Psi^1_{DF}(T) \rangle| =e^{-{\rm Im} \Phi^{E}_{12}} < 1,$
leading to an attenuation of the interferometer fringe pattern. In this case, the full quantum state $|\psi(T)\rangle$ given by (\ref{psi}) is entangled, 
indicating the transfer of which-path information on the atomic motion to the environment. The resulting decoherence  
has been theoretically studied \cite{CasimirDecoherence}  and measured \cite{Sonnentag07} for charged particles close to a material surface. 
Here we focus on the complementary effect that is also present in the general formula (\ref{eq:environment quantum states product}) for the complex 
phase $\Phi_{12}.$ In addition to the loss of fringe visibility, the coupling with the dipole and EM field dofs also leads to a displacement of the 
interference fringes, corresponding to the real part ${\rm Re}\,\Phi_{12}, $ which we analyze in more detail in the remaining part of this paper.

\subsection{Diagrammatic expansion of the environment-induced phase}
As in the previous section, 
we follow a linear response approach and treat the dipolar coupling as a small perturbation. 
Thus, we perform a diagrammatic expansion of the time-ordered (and anti-time-ordered) exponentials appearing in the the formula (\ref{eq:environment quantum states product}) for the environment-induced complex phase
 $\Phi_{12}$ . 
 We focus on the lowest-order diagrams yielding a finite phase. Special care is required, since the dipolar coupling Hamiltonians $\hat{\widetilde{H}}_{AF}(r_k(t))$~\eqref{eq:interaction Hamiltonian Heisenberg} taken at different times do not commute. We calculate 
 $\Phi_{12}$
  to first order in the atomic polarizability, allowing  us to approximate 
 $e^{i \Phi_{12}} \simeq 1+i \Phi_{12} .$ This is a valid approximation as long as the distance between the atom and the plate is much larger than the atomic size (this assumption also justifies the electric dipole approximation). 

It follows from (\ref{eq:environment quantum states product})  that first-order diagrams are proportional to ($\langle ... \rangle_0$ denoting the average over
 the intial environment state $| \Psi_{DF}(0) \rangle$)
\begin{equation}
\pm \frac {i} {\hbar}  \int_0^T dt \,  \langle\, \hat{\mathbf{d}}(t) \cdot \hat{\mathbf{E}}({\bf r}_k(t)) \,\rangle_0.
\end{equation}
and as a consequence vanish since the the atom has no permanent dipole moment. 
 
 Thus, we focus on  second-order diagrams, which are quadratic in the EM field and dipole operators. There are two different ways to build second-order diagrams  from Eq.~(\ref{eq:environment quantum states product}): one can either take two interactions pertaining to the same time-ordered (or anti-time-ordered) exponential, or one may take one interaction from each exponential. 
 Diagrams of 
 the first kind correspond to a sequence of interactions along the same path, and are  referred to  as ``single-path'' (SP) diagrams.
 Diagrams of 
  the second kind  involve simultaneously two distinct paths, and are thus called ``double-path'' (DP) diagrams. The two contributions sum up to give the complex environment-induced phase $\Phi_{12}= \Phi^{\rm SP}_{12}+\Phi^{\rm DP}_{12} $.\\

\subsubsection{Phase contribution of local single-path diagrams}

We consider first the two possible SP diagrams, beginning with the diagram arising from the time-ordered exponential evaluated along the path $1$ in 
the r.-h.-s. of (\ref{eq:environment quantum states product}), whose
contribution  reads:
\begin{equation}
 \Phi^{\rm{SP}}_{1 } = \!  \frac {i} {\hbar^2}  \! \int_0^T dt \! \int_0^t dt' \! \sum_{i,j}\langle \:  \hat{d}_i(t) \hat{d}_j(t') \hat{E}_i(r_1(t)) \hat{E}_j(r_1(t'))  \: \rangle_0 
\end{equation}
where we  sum over the  Cartesian indices $i,j=1,2,3$.  
 In order to express the phase $\Phi^{\rm{SP}}_{1 }$ in terms of dipole and electric field Green's functions 
(\ref{Hadamard},\ref{eq:retarded Green functions}), we write 
the product of dipole (or electric field) operators at distinct times (or space-time points) as the half sum of their commutator and anti-commutator. As in Sec.~II,  these contributions can be expressed in terms of the scalar dipole $g^{R(H)}_{\hat{d}}(t,t')$ and the trace of  the
 electric field Green's function $\mathcal{G}^{R(H)}_{\hat{\bf E}}(x;x').$ For the latter we take only the scattering contribution 
 $\mathcal{G}^{R(H),S}_{\hat{\bf E}}(x;x')$ [see Eq.~(\ref{0S})] and 
 then find that ${\rm Re}\, \Phi^{\rm{SP}}_1 = \varphi^{\rm loc}_1$ is precisely
  the local phase~\eqref{eq:local phase general expression narrow wavepackets} obtained in Sec.~\ref{sec:local dynamical Casimir phases}
  for point-like wave-packets. 

An analogous SP diagram comes from the anti-time ordered exponential along path $2$ in the r.-h.-s. of (\ref{eq:environment quantum states product}), 
yielding a similar contribution $\Phi^{\rm{SP}}_{2}$ to the complex phase. The reversed time-ordering leads to an additional
 minus sign in front of each retarded dipole and electric field Green's functions appearing in the expression for the complex phase. 
 Since  ${\rm Re}\,\Phi^{\rm{SP}}_{2}$ contains an odd number of  retarded Green's functions, we find 
  ${\rm Re}\, \Phi^{\rm{SP}}_2 = -\varphi^{\rm loc}_2$
 with the local phase
 $\varphi^{\rm loc}_2$ given again by
  ~\eqref{eq:local phase general expression narrow wavepackets}. 
  Thus, the total contribution of single-path diagrams has a real part 
 \begin{equation}
  \label{eq:phase SP}
{\rm Re}\,\Phi_{12}^{\rm SP} = \varphi^{\rm loc}_1-\varphi^{\rm loc}_2 
\end{equation}
  Since ${\rm Re}\,\Phi_{12}$ represents the phase coherence of path 1 with respect to path 2, it must be anti-symmetric with respect to
   the interchange of the two paths. This property is clearly satisfied by the local contribution (\ref{eq:phase SP}), and will also hold for 
   the non-local double-path contribution discussed in the following. On the other hand, the imaginary part ${\rm Im}\,\Phi_{12},$
   representing decoherence, 
    must be symmetric 
   with respect to the interchange, with both local path contributions being positive and thus leading to an attenuation of fringe pattern. This property is also
   satisfied
   by the result derived from (\ref{eq:environment quantum states product}) since ${\rm Im}\,\Phi_{12}$ contains an even number of retarded Green's functions.

 In short,
 the local approach developed in Section~\ref{sec:local dynamical Casimir phases} provides the correct expressions for the real part of the 
 single-path contributions to the complex phase $\Phi_{12}.$
  However, it is unable to yield even the single-path contributions to the imaginary part of  $\Phi_{12},$
  which represents the decoherence effect. 
  More importantly, 
 the local theory also misses all double-path phase contributions, which we  discuss in the remaining part of this section.

\subsubsection{Phase contribution of the non-local double-path diagram}

We investigate here the double-path diagram, which involve a product of linear terms issued from both the time-ordered and anti-time-ordered exponentials
in the r.-h.-s. of  (\ref{eq:environment quantum states product}):
\begin{eqnarray}
i \Phi^{\rm{DP}}_{12 } & = & \!  \Biggl\langle \:\sum_{i,j} \left( \frac {i} {\hbar}  \! \int_0^T dt'  \hat{d}_i(t') \hat{E}_i(r_2(t')) \right) \nonumber \\
 & \: &  \times \left( \frac {-i} {\hbar}  \int_0^T dt  \hat{d}_j(t) \hat{E}_j(r_1(t)) \right)  \:  \: \Biggr\rangle_0
 \label{DP_pre}
\end{eqnarray}
As previously, we express the product of two dipole and EM field operators as the half sum of their commutators and anti-commutators.
After summing  over the Cartesian indices $i,j$ and discarding the contributions from the free-space electric field Green's functions, we find for the real part
$\phi^{\rm{DP}}_{12 }   \equiv {\rm Re}\,\Phi^{\rm{DP}}_{12 }$
\begin{widetext}
\begin{eqnarray}
\nonumber
\phi^{\rm{DP}}_{12 } \! & \! = \! & \! \frac {1} {4} \! \! \iint_0^T \!  dt' dt \left[ \frac {} {} g_{\hat{d}}^H(t,t') \left( \mathcal{G}_{\hat{\mathbf{E}}}^{R, S} (r_1(t),r_2(t')) - \mathcal{G}_{\hat{\mathbf{E}}}^{R, S} (r_2(t),r_1(t'))  \right) +  g_{\hat{d}}^{R}(t,t') \left( \mathcal{G}_{\hat{\mathbf{E}}}^{H, S} (r_1(t),r_2(t')) - 
 \mathcal{G}_{\hat{\mathbf{E}}}^{H, S} (r_2(t),r_1(t')) \right) \right]\\
 \label{eq:double path phase}
\end{eqnarray}
\end{widetext}
As required for consistency, the r.-h.-s. of  (\ref{eq:double path phase}) is anti-symmetrical under the interchange 
of the two paths, since $\phi^{\rm{DP}}_{12 }$ represents a contribution to the  relative phase of path 1 with respect to path 2. Remarkably, 
this relative phase contribution depends simultaneously on the two distinct paths of the atom interferometer and cannot be split into separate contributions from 
paths 1 and 2. 

The non-negligible contribution to the non-local phase $\phi^{\rm{DP}}_{12 }$ actually comes entirely from the term proportional to 
$ g_{\hat{d}}^H(t,t')$ in Eq.~(\ref{eq:double path phase}), which accounts for the long-lived atomic dipole fluctuations. 
 Eq.~(\ref{eq:double path phase}) shows that the non-local phase results from the  asymmetry between 
 the cross self-interactions involving different wave-packets --- the fluctuating dipole interacting with the electric field sourced by itself 
 at a different location \cite{DoublePath}.

\section{DYNAMICAL CASIMIR PHASES FOR  FINITE-SIZE WAVE-PACKETS}
\label{sec:finite width}

 The previous derivation of the dynamical Casimir phases for point-like atomic wave-packets highlighted the basic physical mechanisms behind the appearance of a non-local double-path Casimir phase. However, 
 usual experimental conditions in Casimir interferometry~\cite{Cronin04,CroninVigue,Lepoutre11} do not match 
 this assumption, since the 
 width of the atomic wave-packets are of the same order of the atom-surface distances. 
 In this section,  we  present a derivation of the dynamical local and non local Casimir phases for finite-width wave-packets. 

As in the  previous section, we consider the interaction picture. However, we no longer consider the interaction Hamiltonian as parametrized by well-defined atomic trajectories. Instead, we now evolve the interaction Hamiltonian with respect to the external atomic dofs associated to the Hamiltonian $\hat{H}_E$, i.e. the time-dependent interaction Hamiltonian can be expressed as a function of the free-evolving dipole $\hat{\mathbf{d}}(t)$, free-evolving electric field $\hat{\mathbf{E}} ( \mathbf{r},t )$ and initial time position operator $\hat{\mathbf{r}}_a$ as $\hat{\widetilde{H}}_{AF}(t)=e^{  \frac {i} {\hbar}   \hat{H}_E t } \left[-\hat{\mathbf{d}}(t) \cdot \hat{\mathbf{E}} ( \hat{\mathbf{r}}_a,t  )  \right] e^{  -\frac {i} {\hbar}   \hat{H}_E t }  $. 
Again, we consider the coherence of the reduced density matrix~\eqref{eq:definition coherence rho12} $\rho_{12}(\mathbf{r},\mathbf{r}',t)$ between the two wave-packets $\psi_E^1(\mathbf{r},t)$ and $\psi_E^2(\mathbf{r}',t)$, related to the free-evolving density matrix coherence
 $\rho^0_{12}(\mathbf{r},\mathbf{r}';T)=\frac12\psi_E^1(\mathbf{r},t)\psi_E^{2*}(\mathbf{r}',t)$ by $\rho_{12}(\mathbf{r},\mathbf{r}';T)=\rho^0_{12}(\mathbf{r},\mathbf{r}';T)e^{i \phi_{12} (\mathbf{r},\mathbf{r}',T)}$. For a small interaction phase $\phi_{12} (\mathbf{r},\mathbf{r}';T)$, a first-order Taylor expansion yields $ \phi_{12} (\mathbf{r},\mathbf{r}';T) \simeq (-i) \delta \rho_{12}(\mathbf{r},\mathbf{r}';T) / \rho^0_{12}(\mathbf{r},\mathbf{r}';t)$. We have introduced the difference between the free and interacting density matrix coherences $\delta \rho_{12}(\mathbf{r},\mathbf{r}';T)=\rho_{12}(\mathbf{r},\mathbf{r}';T)-\rho^0_{12}(\mathbf{r},\mathbf{r}';T)$, determined below in terms of second-order dipolar interaction diagrams. 
 We also define the average interaction phase coherence 
 $\phi_{12} (T)\equiv \iint d^3 \mathbf{r} d^3 \mathbf{r}' |\psi_E^{1}(\mathbf{r},T)|^2 |\psi_E^{2}(\mathbf{r}',T)|^2 \phi_{12} (\mathbf{r},\mathbf{r}';T)$, equivalently expressed as
\begin{equation}
\label{eq:definition average phase}
\phi_{12} (T)= -2i \int d^3 \mathbf{r} \int d^3 \mathbf{r}'   \psi_E^{1*}(\mathbf{r},T) \psi_E^{2}(\mathbf{r}',T) \;\delta\rho_{12}(\mathbf{r},\mathbf{r}';T).
\end{equation}
 At the time $T$, the reduced density matrix can be formally expressed as
\begin{eqnarray}
\nonumber
\rho_{12}(\mathbf{r},\mathbf{r}';T) & = & \frac12\langle \psi_{DF}(0) | \otimes \langle \psi_E^2(0) | \widetilde{\mathcal{T}} \left[ e^{  \frac {i} {\hbar} \int_0^T dt  \hat{\widetilde{H}}_{ AF}(t) } \right]  \\
& \: & \times \left( e^{ \frac i \hbar H_E T} | \mathbf{r}'\rangle \langle \mathbf{r} |  e^{- \frac i \hbar H_E T} \otimes {\bf 1}_{DF}\right) 
\label{eq:atomic density wide nonperturbative1}
 \\
& \: & \times \mathcal{T} \left[ e^{ - \frac {i} {\hbar} \int_0^T dt'  \hat{\widetilde{H}}_{ AF}(t') } \right]|\psi_{E}^1(0) \rangle \otimes |\psi_{DF}(0) \rangle  \nonumber
\end{eqnarray}

Let us first investigate the SP paths terms, which correspond to contributions to 
 $\delta \rho_{12}(\mathbf{r},\mathbf{r}',T)$ arising from quadratic terms issued from the same time-ordered (or anti-time ordered) exponential. One considers without loss of generality the SP phase associated with path $1$, which yields the contribution:
  \begin{eqnarray}
 & &  \delta\rho_{12}^{\rm SP1}(\mathbf{r}_1,\mathbf{r}_2;T)  =   
  \frac {i} {2\hbar^2} \psi_E^{2*}(\mathbf{r}_2,T) \sum_{i,j=1}^3 \int_0^T dt \int_0^{t} dt' \nonumber \\
   & \times &  \int d^3\mathbf{r}\int d^3\mathbf{r}' K(\mathbf{r}_1,T;\mathbf{r},t)  K(\mathbf{r},t;\mathbf{r}',t') \psi_E^1( \hat{\mathbf{r}}',t') \nonumber \\
   & \times & \langle \widetilde{\psi}_{DF}(0) | \hat{d}_i(t)  \hat{E}_i ( \mathbf{r},t  ) \hat{d}_j(t')  \hat{E}_j ( \mathbf{r}',t'  )  |\widetilde{\psi}_{DF}(0) \rangle \nonumber
   \end{eqnarray}
   When taking the average~\eqref{eq:definition average phase} of $\delta\rho_{12}^{\rm SP1}$, one recognizes an integral involving the external atomic propagator~\eqref{eq:definition atomic propagator}, leading to the Casimir phase~\eqref{eq:local phase general wide atomic packets} obtained previously with the local theory. 
   
   On the other hand, one derives the DP phase from Eq.~\eqref{eq:atomic density wide nonperturbative1} by considering the diagrams composed of linear terms issued from both the time-ordered and anti-time ordered exponentials:
   \begin{eqnarray}
   \label{eq:atomic probability function4}
   & &   \delta\rho_{12}^{\rm DP}(\mathbf{r},\mathbf{r}';T)  = \frac12 \sum_{i,j=1}^3 
    \int d^3 \mathbf{r}  \int d^3 \mathbf{r}'    \langle \psi_{DF}(0) |  \nonumber \\ & \: &   \left[  
     \frac {i} {\hbar}  \int_0^T dt' \psi_{E}^{2*}(\mathbf{r}',t') 
   K (\mathbf{r}',t'; \mathbf{r}_2,T)  \hat{d}_i(t')  \hat{E}_i ( \mathbf{r}',t'  )    \right]  \nonumber \\
   & \: &     \left[ - \frac {i} {\hbar} \int_0^T dt   \hat{d}_j(t) \hat{E}_j ( \mathbf{r},t  ) 
   K (\mathbf{r}_1,T;\mathbf{r},t) \psi_{E}^1(\mathbf{r},t)  \right] |\psi_{DF}(0) \rangle. \nonumber 
   \end{eqnarray}
The averaging procedure~\eqref{eq:definition average phase} yields a double-path phase which depends simultaneously on the histories of the two wave-functions corresponding to each interferometer arm. As in Section~\ref{sec:Feynman diagrams}, we express the bilinear averages of the dipole and field operators in terms of Hadamard and retarded Green's functions:
\begin{eqnarray}
\label{eq:DP phase wide atomic packet general}
\phi^{\rm{DP}}_{12 }(T) & = & \frac {1} {4}  \iint_0^T  dt dt'    \iint d^3 \mathbf{r} d^3 \mathbf{r}'    \left| \psi_{E}^1(\mathbf{r},t) \right|^2 |\psi_{E}^{2}(\mathbf{r}',t')|^2 \nonumber \\ 
&\times & \left[ \frac {} {} g_{\hat{d}}^H(t,t') \left( \mathcal{G}_{\hat{\mathbf{E}}}^{R, S} (\mathbf{r},t;\mathbf{r}',t' ) -  \mathcal{G}_{\hat{\mathbf{E}}}^{R, S} (\mathbf{r}',t';\mathbf{r},t )    \right) \right. \nonumber \\
& + & \left.  g_{\hat{d}}^{R}(t,t') \left( \mathcal{G}_{\hat{\mathbf{E}}}^{H, S} (\mathbf{r},t;\mathbf{r}',t' )  - 
 \mathcal{G}_{\hat{\mathbf{E}}}^{H, S} (\mathbf{r}',t;\mathbf{r},t' ) \right) \frac {} {} \right] \nonumber \\
\end{eqnarray}

If one considers that the electric field Green's functions are uniform over the width of atomic wave-packets, one obviously retrieves the nonlocal DP phase~\eqref{eq:double path phase} of Section~\ref{sec:Feynman diagrams} obtained in the narrow atomic wave-packet limit. 
In order to highlight the dependence of the  DP phase on the dynamical atomic motion,
 we Taylor expand the advanced time wave-function 
 $ | \psi_{E}^k(\mathbf{r},t) |^2 \simeq | \psi_{E}^k(\mathbf{r},t') |^2 + \frac {\partial} {\partial t} | \psi_{E}^k(\mathbf{r},t') |^2\tau $
 in Eq.~(\ref{eq:DP phase wide atomic packet general}). 
 This is an excellent approximation since the time $\tau=|{\bf r}-{\bf r}'|/c$ corresponds to the light propagation between the dipole and its image, and is thus extremely short compared to the typical time scale of the external atomic motion. 
 As before, we assume a stationary regime and write $g^{R,H}_{\hat{d}}(\tau)\equiv 
 g^{R,H}_{\hat{d}}(t'+\tau,t')$.
 Using the conservation of the atomic probability, one can express the DP phase (\ref{eq:DP phase wide atomic packet general}) in terms of the probability current $\mathbf{j}^k(\mathbf{r},t) = \mbox{Re} \left[ \psi_E^{k*}(\mathbf{r},t)  \frac {\hbar} {i m}  \nabla  \psi_E^k(\mathbf{r},t) \right]$:
\begin{eqnarray}
\nonumber
& &  \phi^{\rm{DP}}_{12 }(T) =  \frac {1} {4 } \sum_{i=1}^3 \int_0^T  dt' \int_0^{T-t'} d \tau    \iint d^3 \mathbf{r} d^3 \mathbf{r}'  \\
  & \times &    \left( \frac {} {} j_i^1(\mathbf{r},t')  |  \psi_{E}^2(\mathbf{r}',t') |^2 
 -  j_i^2(\mathbf{r},t') |  \psi_{E}^1(\mathbf{r}',t') |^2 \frac {} {}  \right) \: \tau   \nonumber \\
 & \times &  \left( \frac {} {}  g_{\hat{d}}^H(\tau)  \frac {\partial \mathcal{G}_{\hat{\mathbf{E}}}^{R, S} (\mathbf{r},t'+\tau;\mathbf{r}',t' )} {\partial r_i}  \right.  \nonumber \\
 & \: & \left. \frac {} {}+ g_{\hat{d}}^{R}(\tau)  \frac {\partial \mathcal{G}_{\hat{\mathbf{E}}}^{H, S} (\mathbf{r},t'+\tau;\mathbf{r}',t' )} {\partial r_i}    \right).    \label{eq:DP phase current probability}
  \end{eqnarray}
  The 
 non-local DP phase 
 is thus a dynamical phase correction, with the 
 current density giving the 
 probability density evolution during the very short electromagnetic propagation time $\tau.$
 In the next section, we investigate in greater detail the phases acquired by wide wave-packets flying close to a planar perfectly-reflecting surface.

\section{NON-LOCAL DYNAMICAL CORRECTIONS TO THE VAN DER WAALS PHASE FOR A PLANE SURFACE}

\label{section:relativistic expansion}

In this section, we derive explicit results for the non-local dynamical contributions to the Casimir phase, working at the leading order in  $v/c$ 
($v$ denotes the magnitude of the atomic center-of-mass velocity). Starting from the general results of Sec.~\ref{sec:finite width}, we describe such corrections for wide atomic packets interacting with a perfectly-reflecting planar surface, located at $z=0.$ Moreover,
  we shall consider specifically the short-distance van der Waals (vdW) regime probed by
  the experiments \cite{Cronin04,CroninVigue,Lepoutre11},
  which corresponds to a stronger atom-surface interaction (thus yielding larger dynamical phase corrections)
   than the long-distance Casimir-Polder limit. As discussed in Section~\ref{sec:local dynamical Casimir phases}, at these distances the dominant  dynamical vdW phase contributions come from the electric field response to dipole fluctuations. 
   The experiments were performed for  wide atomic wave-packets filling in the gap between the central trajectory and the conducting plate  \cite{Cronin04,CroninVigue,Lepoutre11}. In this case, we show here that 
 the non-local DP phase is enhanced with respect to the result for
 point-like packets~\cite{DoublePath}  by a logarithmic factor.

  We take a Mach-Zehnder atom interferometer in the half-space $z>0$ 
 close to the material surface at $z=0$ as illustrated by Fig.~\ref{fig:atom interferometer}. 
The two central atomic trajectories share the same velocity component parallel to the plate, but have arbitrary normal velocities: 
\begin{equation}
\label{eq:atomic trajectories}
\mathbf{r}_k(t)= \mathbf{r}_{0 / \! \! /} (t)+ z_k(t) \,\hat{\mathbf{z}}, \quad k=1,2.
\end{equation}
The results to follow can be extended to discuss dynamical vdW phase corrections resulting from atomic interactions with a grating as
 in Refs.~\cite{Cronin04,CroninVigue,Lepoutre11}.

\subsection{Electric field and dipole Green's functions}

It is necessary, at this stage, to have at hand explicit expressions for the dipole and electric field Green's functions. As discussed in Section II, the electric field Green's functions is decomposed as the sum of  free and   scattering contributions. Only the latter is relevant for the derivation of the Casimir phases induced by the surface. 
   We first derive the field Green's functions in Fourier space by writing the electric field operator as a sum over normal modes, taking due account 
   of the perfectly-reflecting surface at $z=0.$  We then 
   derive both the known result for the free-space Green's function~\cite{Heitler} as well as the scattering contribution
   \begin{eqnarray}
   \label{G_sca}
    \mathcal{G}_{\hat{\mathbf{E}}}^{R, S}(x,x')   & = 
    & \frac {\theta(\tau)} {2 \pi\epsilon_0} \frac {\partial^2} {\partial z \partial z'} \left( \frac {\delta ( \tau-|\mathbf{r}-\mathbf{r}'_{\rm I}|/c )} {|\mathbf{r}-\mathbf{r}'_{\rm I}|}  \right) 
   \end{eqnarray} 
   As expected $ \mathcal{G}_{\hat{\mathbf{E}}}^{R, S}(x,x') $ 
   depends on the time difference $\tau=t-t'$ only and not on the individual times. 
   It is written in terms of the
 propagation distance $|\mathbf{r}-\mathbf{r}'_{\rm I}|$ between the point $\bf{r}$ and the 
    image  $\mathbf{r}'_{\rm I}=(x',y',-z')$  of the source point ${\bf r}'=(x',y',z')$ with respect to the plane surface.
     Assuming the EM field
    to be in thermal equilibrium, the electric field Hadamard Green's function
     $ \mathcal{G}_{\hat{\mathbf{E}}}^{H, S} (x,x')$
     can be obtained from the retarded one thanks to the fluctuation-dissipation theorem.

   In order to obtain the dipole Green's functions, we model the internal atomic degrees of freedom as an harmonic oscillator with a transition
    frequency $\omega_0$ (and wave-length $\lambda_0$) and assume 
   the atom to be in its ground state.
The Hadamard dipole Green's function is then proportional to the static atomic polarizability $\alpha(0):$ 
       \begin{equation}
       \label{eq:dipole Hadamard Green}
          g^{H}_{\hat{d}}(t,t')= \alpha(0)\, \omega_0 \,\cos[\omega_0(t-t')].
          \end{equation}

\subsection{Nonlocal dynamical phases}


We consider the limit of wide atomic packets with a well-defined momentum, which is well-suited to describe the dispersion effects associated to the finite width of the atomic packets propagating nearby the plate. In this limit, one may take the probability current involved in the DP path phase~\eqref{eq:DP phase current probability} as 
$\mathbf{j}^k(\mathbf{r},t) \simeq |\psi_E^{k}(\mathbf{r},t)|^2 \mathbf{v}_k(t)$ where $\mathbf{v}_k(t)={\bf\dot r}_k(t)$ is a classical velocity \cite{RemarkWidePacketApprox}. Since the DP phase depends sharply on the distance between the atoms and the conductor and not on their lateral position above this surface, the extension of the atomic wave-packets in the direction $O_z$ normal to the conducting surface is much more critical than the extension of the atomic packets along the directions $O_x,O_y$ parallel to the conductor. Thus, one can safely use one-dimensional atomic wave-packets $\psi_E^{1,2}(z,t)$ in order to model dispersion effects in the nonlocal DP phase acquired by wide atomic beams.

We first model the atomic wave-functions by a step-wise distribution centered on the classical atomic trajectories of time-independent width, i.e. we take $|\psi_{k}^E(z,t)|^2=1/w$ for $z_k(t)-w/2 < z < z_k(t)+w/2$ and zero for $|z-z_k(t)| \geq w/2$ -- with a width $w$ such that $w \leq 2 z_0$ where $z_0=z_1(0)=z_2(0)$ is the initial distance between the atomic wave-packet centers and the plate. Naturally, such description is a simple approximation, and a modelling in terms of Gaussian wave-packets would be more accurate. Nevertheless, this approach should yield the correct qualitative picture and has the advantage of giving analytical expressions regarding the dependence of the DP phase towards the wave-packet width.

We calculate the DP phase in the short-distances vdW regime and 
 take $g_d^H(\tau) \approx g_d^H(0) = \omega_0 \alpha(0)$ [see (\ref{eq:dipole Hadamard Green})]. We consider the linear trajetories~\eqref{eq:atomic trajectories}, and assume that the
   distance between the central trajectory endpoints is much larger than the initial altitude $z_0$, yielding the saturation limit of the DP phase~\cite{DoublePath}. Using the step wave-functions in Eq.\eqref{eq:DP phase current probability}, one
     obtains an expression for the DP phase taking into account the finite atomic packet extension:
\begin{equation}
\phi^{\rm DP}_{12}(z_0,w) = - \frac {3\pi} {\lambda_0} \frac {\alpha(0)} {4\pi \epsilon_0}  \frac{1}{w^2} \log \left(  1 - \frac {w^2} {4 z_0^2} \right)  
\end{equation}
When taking the limit $w \ll z_0$ in this expression, one retrieves the DP phase obtained in~\cite{DoublePath} for classical trajectories. On the other hand, the phase $\phi^{\rm DP}_{12}(z_0,w)$ diverges when the wave-packet width $w$ approaches $2 z_0$, i.e. when the edge of the atomic wave-function becomes close to the plate. This suggests that a greater care is needed to evaluate this phase when considering atomic wave-functions which do not vanish at the plate boundary, where the vdW potential becomes infinite. 

Indeed, the divergence above is a consequence of our perturbative approach, jointly with the 
the small phase approximation  $e^{i \phi^{\rm DP}_{12}} \simeq 1+ i\phi^{\rm DP}_{12}$, which obviously breaks down
at the close vicinity of the plate (dispersion interaction models in general are valid only for distances much larger than the atomic length scale). 
 Fortunately, this divergence can be easily cured, since such contributions lead to quickly oscillating complex exponentials which in fact barely affect the average vdW phase~\cite{Cronin04,Lepoutre11}. To make our argument more precise, we reintroduce these  exponentials in our derivation of the average dynamical phase 
 $\Phi^{\rm{DP}}_{12 }(T)$:
 \begin{eqnarray}
 |A| e^{i \Phi^{\rm{DP}}_{12 }(T)} & = &  \int d z^0_1  d z^0_2  |\psi_E^1(z^0_1,0)|^2  |\psi_E^2(z^0_2,0)|^2 \label{eq:nonperturbative DP phase}  \\
 & \: & \qquad \qquad \qquad \qquad \times e^{i \phi_{12}^{\rm DP}(z^0_1,z^0_2,T)}   \nonumber
\end{eqnarray}
with the phase 
\begin{eqnarray}
& & \phi_{12}^{\rm DP}(z^0_1,z^0_2,T)   =   \frac 1 4 \int_0^T dt' \int_0^{T-t'} d\tau g_{\hat d}^H(\tau) \tau  \nonumber  \\
& \: &   \times \left( v_{1 \: z}(t') - v_{2 \: z}(t') \right)  \frac {\partial} {\partial z} \mathcal{G}^R_E \left( \frac {} {} z_1(t') \mathbf{\hat z} ,
t'+\tau;z_2(t')\mathbf{\hat z},t'  \frac {} {} \right) \nonumber
\end{eqnarray}
and $z_k(t')=z^0_k+\int_0^{t'} dt'' v_{k \: z}(t'').$ We have omitted the common displacement of the atomic wave-packets parallel to the plate on both trajectories thanks to the translational invariance of the field Green's function along this direction. Using the vdW regime and the saturation limit, and following Ref.~\cite{DoublePath}, one finds 
\begin{equation}
\label{eq:phi DP origin z01 z02}
\phi_{12}^{\rm DP}(z^0_1,z^0_2,T) = \frac  {3 \pi   } {\lambda_0 } \left(  \frac {\alpha(0)}  {4 \pi \epsilon_0 } \right) (z^0_1+z^0_2)^{-2} 
\end{equation}.

Eqs.(\ref{eq:nonperturbative DP phase},\ref{eq:phi DP origin z01 z02}) are the starting point of the derivation to follow. We consider initial atomic wave-functions filling in the gap between the central atomic position and the material surface, taking again a step wave-function approach with this time $w=2 z_0$. 

Under the above approximations and following the averaging procedure of Refs.~\cite{Cronin04,Lepoutre11}, one derives the average DP phase $ \tan \phi^{\rm DP}_{12}(w)= I_s/ I_c $ with $I_s = (w_c^2/ 2 w^2) \int_{w_c^2 /w^2}^{+\infty} d \phi     \phi^{-2} \sin ( \phi )$ and $I_c = (w_c^2/ 2 w^2) \int_{w_c^2 /w^2}^{+\infty} d \phi     \phi^{-2} \cos ( \phi ).$ We have introduced a critical length scale associated with the DP phase $w_c=[\frac {3  \pi } {\lambda_0} \left(\alpha(0)/(4 \pi \epsilon_0)\right)]^{1/2}.$ The distance $r_{\alpha}=[\alpha(0)/(4 \pi \epsilon_0)]^{1/3}$ 
represents the atomic length scale and is 
of the order of the Angstr\"om. Thus, the length  $w_c = \sqrt{3 \pi}  r_{\alpha} (r_{\alpha}/\lambda_0) ^{1/2}$ is always
 several orders of magnitude smaller than any experimentally achievable atomic packet width $w$. Thus, one may keep only the lowest-order quadratic terms in the small parameter $w_c/w$, taking $I_c \simeq 1$ and
\begin{equation}
\phi^{\rm DP}_{12}(w) = \frac  {3  \pi } {\lambda_0} \left(\frac {\alpha(0)} {4 \pi \epsilon_0}\right) \frac {1} {w^2} \ln \left( \frac {w} {w_c} \right) + O\left(\frac {w_c^4} {w^4}\right)
\end{equation}
 A comparison with the
results for point-like packets  following identical central trajectories~\cite{DoublePath} 
shows that wide atomic beams experience an enhancement of the DP phase by a factor
 $ \ln \left( \frac {w} {w_c} \right).$ 
 Considering $\:^{87} \mbox{Rb}$ atoms and a wave-packet width $w  = 40 \: {\rm nm}$ (and thus $z_0  = w_0/2 = 20 \: {\rm nm}$) compatible with the parameters used in the Casimir experiments~\cite{Cronin04,CroninVigue,Lepoutre11} for the wave-packets, one obtains a DP phase $\phi^{\rm DP}_{12 \: w} \simeq 3 \times 10^{-6} \: {\rm rad},$ 
 corresponding to an enhancement of 
 roughly one order of magnitude.

\section{CONCLUSION}
\label{section:conclusion}

Using standard perturbation theory, 
we have addressed dynamical corrections, arising from the external motion, to the Casimir phase acquired by neutral atoms interacting with a material surface. A careful description of retardation effects, combined with the atomic motion, reveals the appearance of a non-local atomic phase coherence, which involves  simultaneously a pair of atomic paths instead of a single atomic trajectory as usual in atom optics.  

By construction, the non-local  phase for a given pair of paths must be anti-symmetric with respect to the interchange of the two paths in the pair. 
In fact, it results 
from the difference between the 
 EM propagation distances from one path to the other one after one reflection at the surface.
 Thus, it vanishes when the two path motions with respect to the plate are symmetrical (as for instance in the case of trajectories parallel to plate). 
 In other words, the symmetry between the two paths is broken by the velocity components normal to the surface and the non-local phase is proportional 
 the difference between the two velocity components of a given pair.

In a previous work~\cite{DoublePath}, 
   we had obtained a preliminary estimation of the non-local double-path
    phase for point-like atomic wave-packets
    using an independent and less intuitive method based on the influence functional.
    Here we have obtained  these dynamical Casimir phases by keeping track of the quantum state of the environment -- the EM field and the atomic dipole degrees of freedom. This treatment provides us with an interesting open-system interpretation of this double-path atomic phase coherence, by showing that it results from a non-local disturbance of the environment by a coherent superposition of external atomic waves propagating across two distinct atomic paths. The approach developed here also corresponds to more realistic experimental conditions, since it takes into account the atomic dispersion in position around the central path, which is relevant for the estimation of the vdW phase~\cite{CroninVigue}. The corresponding general expressions, written in terms of Green's functions for the field and  atomic internal dofs, and of the atomic probability current and wave-functions,
    are in principle valid for arbitrary geometries and non-equilibrium conditions. We have also derived explicit analytical results
         for a perfectly-reflecting planar surface in the short-distance regime. In this regime, our treatment reveals a significant enhancement of the non-local DP phase acquired by wide atomic packets with respect to our previous estimation based simply on classical atomic trajectories.

Both the local and non-local dynamical atomic Casimir phases are first-order relativistic corrections arising from the external atomic motion, and thus of similar magnitude. This shows that the relativistic corrections to the Casimir phase are intrinsically non-local.


\begin{acknowledgments}

The authors are grateful to Reinaldo de Melo e Souza for stimulating discussions. 
This work was partially funded  by CNRS (France), CNPq, FAPERJ and CAPES (Brazil).
\end{acknowledgments}

\appendix*

\section{QUASI-STATIC LIMIT OF THE LOCAL  ATOMIC PHASE}

Here, we assume that the field is in thermal equilibrium, and we consider the regime of long atom-surface interaction times, namely we take an atomic time-of-flight $T$ above the conductor much larger than the atomic dipole or field correlation time scales. In this regime, we show that
the non-relativistic contribution to the local Casimir phase of Section~\ref{sec:local dynamical Casimir phases} reduces to the standard phase arising from a dispersive (Casimir) potential. Taking the quasi-static limit of Eq.~\eqref{eq:local phase general wide atomic packets}, one obtains
\begin{eqnarray}
\label{phi_local_qs}
\varphi^{\rm loc}_k & \approx & \! \frac 1 4 \int_{0}^{T} \! d t'  \int \! d^3\mathbf{r} |\psi_E^k(\mathbf{r},t)|^2  \\
& & \times \int_0^t  d\tau \left[ \frac {} {} g_{\hat{d}}^H(\tau) \mathcal{G}_{\hat{\mathbf{E}}}^{R,S}(\mathbf{r},\mathbf{r};\tau)
 +    g_{\hat{d}}^R(\tau)  \:   \mathcal{G}_{\hat{\mathbf{E}}}^{H,S}(\mathbf{r},\mathbf{r};\tau) \frac {} {}  \right] \nonumber
\end{eqnarray}
We have assumed 
that the dipole and field fluctuations are stationary
 in order to write $g_{\hat{d}}^{R(H)}(\tau)  \equiv g_{\hat{d}}^{R(H)}(t+\tau,t)$ and $\mathcal{G}_{\hat{\mathbf{E}}}^{R(H),S}(\mathbf{r},\mathbf{r};\tau) \equiv \mathcal{G}_{\hat{\mathbf{E}}}^{R(H),S}(\mathbf{r},t+\tau;\mathbf{r},t) $.

In the equation above, we focus on the integral over the delay $\tau$, whose bounds can be extended to infinity in the regime of large atom-surface interaction times. Using the Parseval-Plancherel relation, we express the local phase in the Fourier domain as follows
 \begin{eqnarray}
 \label{eq:nonrelativstic limit3}
& & \varphi^{\rm    loc }_k  \approx  \frac {1} {8 \pi}   \int_{0}^{T} d t  \int \! d^3\mathbf{r} |\psi_E^k(\mathbf{r},t)|^2 \\
 & \times &  \int d \omega \left( \! g_{\hat{d}}^R(\omega)  \:   \mathcal{G}_{\hat{\mathbf{E}}}^{H,S *}(\mathbf{r},\mathbf{r};\omega) +    \mathcal{G}_{\hat{\mathbf{E}}}^{R,S}(\mathbf{r},\mathbf{r}; \omega)  g_{\hat{d}}^{H *}(\omega)  \right)  \nonumber 
 \end{eqnarray}
The Fourier transform of the Green's function is defined as:
\begin{eqnarray}
 g^{R(H)}_{\hat{d}}(\omega)  =  \int_{-\infty}^{+\infty}  d\tau  g_{\hat{d}}^{R(H)}(\tau) e^{i \omega \tau} \nonumber 
\end{eqnarray}
and likewise for 
 $\mathcal{G}^{R(H),S}_{\hat{\mathbf{E}}}(\mathbf{r},\mathbf{r};\omega).$

Our next step is to express the dispersive potential as a similar frequency integral. 
We assume that the electric field and dipole dofs are at thermal equilibrum at temperature $\Theta.$
One starts with the general expression derived in Ref.~\cite{WylieSipe}:
\begin{equation}
\label{eq:potentialCPgeneralexpression}
V_{\rm{Cas}}(\mathbf{r}) \! = \! - \frac {\hbar} {2 \pi} \! \int_{0}^{+\infty} \! \! \! \! d \omega \, \coth \left( \frac {\hbar \omega} {2 k_B \Theta} \right)
   {\rm{Im}} \left[ \! g^R_{\hat{d}}(\omega)  \mathcal{G}^{R,S}_{\hat{\mathbf{E}}}(\mathbf{r},\mathbf{r};\omega) \! \right]
\end{equation}
where $k_B$ is the Boltzmann constant. 
In order to cast (\ref{eq:potentialCPgeneralexpression}) in the form of Eq.~\eqref{eq:nonrelativstic limit3}, we use the fluctuation-dissipation theorem (FDT): 
\begin{eqnarray}
g^H_{\hat{d}}(\omega) \! & \! = \!  & \! 2  \coth \left( \frac {\hbar \omega} {2 k_B \Theta} \right) \mbox{Im} \left[ g^R_{\hat{d}}(\omega)   \right] \\
 G^{H,S}_{\hat{\mathbf{E}}}(\mathbf{r},\mathbf{r};\omega) \! & \! = \! & \! 2  \coth \left( \frac { \hbar \omega} {2 k_B \Theta} \right) \mbox{Im} \left[ G^{R,S}_{\hat{\mathbf{E}}}(\mathbf{r},\mathbf{r};\omega)  \right] \nonumber
\end{eqnarray}
Using these relations, we rewrite \eqref{eq:potentialCPgeneralexpression} as
\begin{eqnarray}
\label{eq:potential Fourier expression2}
V_{\rm{Cas}}(\mathbf{r}) \! & \! = \! & \! - \frac {\hbar} {4 \pi} \! \int_{0}^{+\infty} \! \! d \omega  \left\{ G^H_{\hat{d}}(\omega)   \mbox{Re} \left[  \mathcal{G}^{R,S}_{\hat{\mathbf{E}}}(\mathbf{r},\mathbf{r};\omega) \right] \nonumber \right. \\
&\: & \qquad \qquad \: \left. + \mbox{Re} \left[  g^R_{\hat{d}}(\omega) \right]  \mathcal{G}^{H,S}_{\hat{\mathbf{E}}}(\mathbf{r},\mathbf{r};\omega) \right\}
\end{eqnarray}
Then, we use the parity of the Green's functions with respect to the frequency $\omega$ in order to extend the lower bound of the integral
in (\ref{eq:potential Fourier expression2})
 to $-\infty$. Note that  $g^{(R,H)}_{\hat{d}}(-\omega)=g^{(R,H)*}_{\hat{d}}(\omega)$ since the Green's functions $g^{(R,H)}_{\hat{d}}(t,t')$ are real. In addition, the FDT shows that $g^{H}_{\hat{d}}(\omega)$ is real. Similar relations hold for the electric 
field Green's functions $\mathcal{G}^{(R,H),S}_{\hat{\mathbf{E}}}(\mathbf{r},\mathbf{r};\omega)$. One then derives
\begin{eqnarray}
\label{eq:potential Fourier expression3}
V_{\rm{Cas}}(\mathbf{r}) \!  & \! = \! & \! - \frac {\hbar} {8 \pi} \! \int \! d \omega  \left( g^H_{\hat{d}}(\omega)   \mbox{Re} \left[\mathcal{G}^{R,S}_{\hat{\mathbf{E}}}(\mathbf{r},\mathbf{r};\omega) \right] \right.  \\ 
& & \qquad \qquad \qquad \left. + \mbox{Re} \left[ g^R_{\hat{d}}(\omega) \right]  \mathcal{G}^{H,S}_{\hat{\mathbf{E}}}(\mathbf{r},\mathbf{r};\omega) \right) \nonumber
\end{eqnarray}
We can add  
 $ g^H_{\hat{d}}(\omega)  \mbox{Im} \left[  \mathcal{G}^{R,S *}_{\hat{\mathbf{E}}} \left(\mathbf{r},\mathbf{r};\omega\right) \right]$ and $ \mbox{Im} \left[g^{R*}_{\hat{d}}(\omega)\right] \mathcal{G}^{H,S}_{\hat{\mathbf{E}}} \left(\mathbf{r},\mathbf{r};\omega\right)$  to the integrand in  (\ref{eq:potential Fourier expression3}) since they are odd functions of
 $\omega:$
\begin{eqnarray}
\label{eq:potential Fourier expression4}
V_{\rm{Cas}}(\mathbf{r}) \! & = & \!  \frac {-\hbar} {8 \pi} \! \int \! d \omega \! \left( \! g^H_{\hat{d}}(\omega)  \mathcal{G}^{R,S *}_{\hat{\mathbf{E}}} \! \left(\mathbf{r},\mathbf{r};\omega\right) \right. \\
& \: & \qquad \qquad \qquad \qquad \left. +   \mathcal{G}^{H,S}_{\hat{\mathbf{E}}}(\mathbf{r},\mathbf{r};\omega) g^{R*}_{\hat{d}} \! (\omega) \! \right)  \nonumber 
\end{eqnarray}

By inspection of Eqs.~\eqref{eq:nonrelativstic limit3} and (\ref{eq:potential Fourier expression4}), we conclude that the
 local Casimir phase in the quasi-static limit takes the standard form~\eqref{eq:nonrelativstic limit 5 wide} of an atomic Casimir phase~\cite{CroninVigue}.


\begin{thebibliography}{25}
\expandafter\ifx\csname
natexlab\endcsname\relax\def\natexlab#1{#1}\fi
\expandafter\ifx\csname bibnamefont\endcsname\relax
  \def\bibnamefont#1{#1}\fi
\expandafter\ifx\csname bibfnamefont\endcsname\relax
  \def\bibfnamefont#1{#1}\fi
\expandafter\ifx\csname citenamefont\endcsname\relax
  \def\citenamefont#1{#1}\fi
\expandafter\ifx\csname url\endcsname\relax
  \def\url#1{\texttt{#1}}\fi
\expandafter\ifx\csname
urlprefix\endcsname\relax\def\urlprefix{URL }\fi
\providecommand{\bibinfo}[2]{#2}
\providecommand{\eprint}[2][]{\url{#2}}

\bibitem{CasimirPolder} H. B. Casimir and  D. Polder, Phys. Rev. {\bf 73}, 360 (1948).

\bibitem{Intravaia} F. Intravaia, C. Henkel and M. Antezza, in {\it Casimir Physics},  edited by D. Dalvit, P. Milonni, D. Roberts and F. da Rosa,
Lecture Notes in Physics No. 834,
 (Springer, Berlin, 2011), Chap. 11, and references therein. 

\bibitem{QuantumFrictionPP} E. V. Teodorovitch,  Proc. R. Soc. A {\bf 362},  71 (1978); Schaich W L and Harris J, J. Phys. F: Met. Phys. {\bf 11} 65 (1981);
 J. B. Pendry,  J. Phys.: Cond.  Matter  {\bf  9},  10301 (1997);  
A. I. Volokitin and B. N. J. Persson,  Phys. Rev. B {\bf 74}, 205413 (2006);
 T. G. Philbin and U. Leonhardt, New J. Phys. {\bf 11}, 033035 (2009);
J. B. Pendry   New J. Phys. {\bf 12}, 033028 (2010);
C. D. Fosco, F. C. Lombardo and F. D. Mazzitelli, Phys. Rev.  D {\bf 84}, 025011 (2011);
 G. Barton,  J. Phys.: Condens. Matter {\bf 23} 355004 (2011).
 

\bibitem{QuantumFrictionAP}  J. F. Annett and  P. M. Echenique, Phys. Rev. B {\bf 34}, 6853  (1986); 
A. I. Volokitin and B. N. J. Persson, Phys. Rev. B {\bf 65}, 115419 (2002);
A. A. Kyasov and G. V. Dedkov, Phys. Solid State {\bf 44},
1809 (2002);
G. Barton, New J. Phys. {\bf 12} 113045 (2010); 
F. Intravaia, R. O. Behunin and D. A. R. Dalvit, arXiv:1308.0712 (2013);
P. W. Milonni, arXiv:1309.1490 (2013). 

\bibitem{Scheel09} S. Scheel  and S. Y. Buhmann,  Phys. Rev. A {\bf 80}, 042902 (2009). 

\bibitem{Cronin09} A. D. Cronin, J. Schmiedmayer and D. E. Pritchard, Rev. Modern Phys. {\bf 81}, 1051 (2009) and references therein.

\bibitem{Kasevich07}  J. M Hogan, D. M. S. Johnson, M. A. Kasevich, in \textit{Proc. Int. School of Physics Enrico Fermi} (2007) and references therein.

\bibitem{Cronin04}  A.D. Cronin, J.D. Perreault, Phys. Rev. A {\bf 70}, 043607 (2004).

\bibitem{CroninVigue} J. D. Perreault and A. D. Cronin, Phys. Rev. Lett.~\textbf{95}, 133201 (2005);  Phys. Rev. A~\textbf{73}, 033610 (2006);
S. Lepoutre, H. Jelassi, V. P. A. Lonij, G. Tr\'enec, M. B\"uchner, A. D. Cronin and J. Vigu\'e, Europhys. Lett.~\textbf{88}, 20002 (2009).

\bibitem{Lepoutre11}
 S.~Lepoutre, V.~P.~A. Lonij, H.~Jelassi, G.~Tr\'enec, M.~B\"uchner, A.~D. Cronin, and J.~Vigu\'e,  Eur. Phys. J. D \textbf{62}, 309 (2011).


\bibitem{FORCAGpapers} P. Wolf, P. Lemonde, A. Lambrecht, S. Bize, A. Landragin, and A. Clairon, Phys. Rev. A~\textbf{75}, 063608 (2007); S. Pelisson, R. Messina, M.-C. Angonin, and P. Wolf, Phys. Rev. A~\textbf{86}, 013614 (2012).

\bibitem{MultiplePathAtomInterferometer}
\bibinfo{author}{\bibfnamefont{M.}~\bibnamefont{Weitz}}, 
  \bibinfo{author}{\bibfnamefont{T.}~\bibnamefont{Heupel}}, \bibnamefont{and}
  \bibinfo{author}{\bibfnamefont{T.~W.} \bibnamefont{H\"ansch}},
  \bibinfo{journal}{Phys. Rev. Lett.} \textbf{\bibinfo{volume}{77}},
  \bibinfo{pages}{2356} (\bibinfo{year}{1996}); \bibinfo{author}{\bibfnamefont{H.}~\bibnamefont{Hinderth\"ur}}~\bibnamefont{et~al.},
    \bibinfo{journal}{Phys. Rev. A} \textbf{\bibinfo{volume}{56}},
    \bibinfo{pages}{2085} (\bibinfo{year}{1997}); \bibinfo{author}{\bibfnamefont{H.}~\bibnamefont{Hinderth\"ur}}~\bibnamefont{et~al.},
  \bibinfo{journal}{Phys. Rev. A} \textbf{\bibinfo{volume}{59}},
  \bibinfo{pages}{2216} (\bibinfo{year}{1999}); F. Impens, C. J. Bord\'e, Phys. Rev. A~\textbf{80} 031602 (2009); M. Robert-de-Saint-Vincent~et al., Europhysics Lett.~\textbf{89} 10002 (2010); F. Impens, F. Pereira dos Santos, and C. J. Bord\'e, New J.  Phys.~\textbf{13}, 065024 (2011).

\bibitem{NonAdditiveCasimir} F. Impens, C. Ccapa Ttira, and P. A. Maia Neto, J. Phys. B: At. Mol. Opt. Phys. {\bf 46},  245503 (2013).

\bibitem{DoublePath} F. Impens, R. O. Behunin, C. Ccapa Ttira, and P. A. Maia Neto, Europhysics Lett.~\textbf{101}, 60006 (2013).

\bibitem{FeynmanVernon} R. P. Feynman and F. L. Vernon, Ann. Phys. (N.Y.) \textbf{24}, 118 (1963).

\bibitem{Ryan10}  R. O. Behunin, and  B.-L. Hu, J. Phys. A: Math. Theor. \textbf{43}, 012001 (2010); Phys. Rev. A \textbf{82}, 022507 (2010).

\bibitem{Ryan11} R.O. Behunin, and B.-L. Hu, Phys. Rev. A~\textbf{84}, 012902 (2011).


\bibitem{CalzettaHu} E. A. Calzetta and B.-L. Hu, \textit{Nonequilibrium Quantum Field Theory}, (Cambridge University Press, Cambridge, UK, 2008). 


\bibitem{SAI90}  A. Stern, Y. Aharonov, and Y. Imry, Phys. Rev. A \textbf{41}, 3436 (1990)

\bibitem{Barone} P. M. V. B.  Barone and A. O. Caldeira, Phys. Rev. A {\bf 43}, 57 (1991).

\bibitem{Hackermueller04} 
L. Hackermueller, K. Hornberger, B. Brezger, A. Zeilinger and M. Arndt, 
Nature \textbf{427}, 711 (2004).

\bibitem{Breuer01} H.-P. Breuer and F. Petruccione, Phys. Rev. A {\bf  63}, 032102 (2001). 

\bibitem{Lamine06}  B. Lamine, R. Herv\'e, A. Lambrecht and S. Reynaud, Phys. Rev. Lett. {\bf 96}, 050405 (2006).

\bibitem{CasimirDecoherence} 
L. H. Ford, Phys. Rev. D {\bf 47}, 5571 (1993); 
J. R. Anglin and W. H. Zurek, 
   in \textit{Dark Matter in Cosmology, Quantum Measurements, Experimental Gravitation}, p.263-270, edited by R. Ansari, Y. Giraud-Heraud and
   J. Van Tran Tranh
      (Editions Frontieres, Gif-sur-Yvette, 1996); 
S.  Scheel and S. Y. Buhmann, 
 Phys. Rev. A \textbf{85}, 030101(R)  (2012).

\bibitem{Sonnentag07}
P. Sonnentag and F. Hasselbach, Phys. Rev. Lett. \textbf{98}, 200402 (2007).

\bibitem{Dalvit00} D. A. R. Dalvit and P. A. Maia Neto, Phys. Rev. Lett. {\bf 84}, 798 (2000);
 P. A. Maia Neto and D. A. R. Dalvit, Phys. Rev. A {\bf 62}, 042103 (2000). 


\bibitem{Mazzitelli03}
F. D. Mazzitelli, J.-P. Paz, and A. Villanueva, Phys. Rev. A \textbf{68}, 062106 (2003).

\bibitem{GeometricPhase} R. S. Whitney, Y. Makhlin, A. Shnirman and Y. Gefen,  Phys. Rev. Lett. \textbf{94}, 070407 (2005); 
F.~C.~Lombardo and P.~I.~Villar, Phys. Rev. A~\textbf{74}, 042311 (2006).

\bibitem{BordeABCD} \bibinfo{author}{\bibfnamefont{C.~J.}~\bibnamefont{Bord\'{e}}},\bibinfo{journal}{C. R. Acad. Sci. Paris} \textbf{\bibinfo{volume}{4}}, \bibinfo{pages}{509} (\bibinfo{year}{2001}{\natexlab{a}});  
   \bibinfo{author}{\bibfnamefont{C.~J.} \bibnamefont{Bord\'{e}}},
  \bibinfo{journal}{Metrologia} \textbf{\bibinfo{volume}{39}},
  \bibinfo{pages}{435} (\bibinfo{year}{2002}{\natexlab{b}}).

\bibitem[{\citenamefont{Bord\'{e}}(1991)}]{BordeHouches}
C. J. Bord\'e, in \textit{Fundamental Systems in Quantum Optics}, Les
Houches Lectures LIII (Elsevier, New York, 1991).

\bibitem{AtomLaserABCD} J.-F. Riou~\bibnamefont{et~al.}, Phys. Rev. A \textbf{77}, 033630 (2008); F.~Impens, Phys. Rev. A~\textbf{80}, 063617 (2009).

\bibitem{DalibardRocCohen}  J. Dalibard,  J. Dupont-Roc, C. Cohen-Tannoudji, J. Phys. (France)~\textbf{43}, 1617 (1982); \textit{ibid}
~\textbf{45}, 637 (1984).

\bibitem{WylieSipe} J. M. Wylie and J. E. Sipe, Phys. Rev. A \textbf{30}, 1185 (1984);  \textbf{32}, 2030 (1985).


\bibitem{Heitler} W. Heitler,  \textit{The Quantum Theory of Radiation}, (Dover, New York, 1954), ch. II; C. Cohen-Tannoudji, J. Dupont-Roc and G. Grynberg
\textit{Photons and Atoms: Introduction to Quantum Electrodynamics}, (Wiley, New York, 1989), ch. III. 

\bibitem{Meschede90} D. Meschede, W. Jhe and E. A. Hinds, Phys. Rev. A {\bf 41}, 1587 (1990).

\bibitem{Mendes} T. N. C. Mendes, C. Farina, J. Phys. A: Math. Gen. {\bf 39}, 6533 (2006). 

\bibitem{Antezza} M. Antezza, L. P. Pitaevskii and S. Stringari, Phys. Rev. Lett. {\bf 95}, 113202 (2005); J. M. Obrecht, R. J. Wild, M. Antezza, L. P. Pitaevskii, S. Stringari and E. A. Cornell, Phys. Rev. Lett. {\bf 98}, 063201 (2007). 

\bibitem{foot_Rontgen} In principle we also need the R\"ontgen interaction term $-{\bf d}\cdot \dot{\bf r}_k \times {\bf B}$~\cite{Scheel09}
in order to have the complete
correction to first-order in $\dot{\bf r}_k.$ However, one can show that the R\"ontgen contribution for short atom-surface distances is much smaller than the dynamical
contribution arising from the 
electric dipolar Hamiltonian calculated here. 


\bibitem{RemarkWidePacketApprox}{Since we  consider below the limit of large trajectory endpoint separation where the DP phase becomes independent of the atomic momentum, the assumption of well-defined atomic momentum is indeed not necessary, even though it permits a formally simpler discussion. Thus, our results would also be valid for atomic packets of intermediate size exhibiting a non-negligible dispersion in atomic momentum.}


\end{thebibliography}
\end{document}